\documentclass[sigconf]{acmart}

\usepackage{booktabs} 
\usepackage{bm} 

\setcopyright{rightsretained}

\newcommand{\s}{\mathbf{s}}

\newcommand{\x}{\mathbf{x}}
\newcommand{\h}{\mathbf{h}}
\newcommand{\Gb}{\mathbf{G}}
\newcommand{\Q}{\mathbf{Q}}
\newcommand{\bk}{^{(a,b)}}
\newcommand{\mub}{\pmb{\mu}}
\newcommand{\Sigmab}{\pmb{\Sigma}}

\DeclareMathOperator*{\argmax}{arg\,max}

\acmDOI{}

\acmISBN{}

\acmPrice{}

\begin{document}

\title{Seasonal Stochastic Blockmodeling for Anomaly Detection in Dynamic Networks}

\author{Jace Robinson}
\affiliation{%
  \institution{Dept. of Computer Science \& Engineering \\ Kno.e.sis Research Center \\Wright State University}
  \city{Dayton} 
  \state{Ohio} 
}
\email{robinson.329@wright.edu}

\author{Derek Doran}
\affiliation{%
  \institution{Dept. of Computer Science \& Engineering \\ Kno.e.sis Research Center \\Wright State University}
  \city{Dayton} 
  \state{Ohio} 
}
\email{derek.doran@wright.edu}

\begin{abstract}
Sociotechnological and geospatial processes exhibit time varying structure that make 
insight discovery challenging. To detect abnormal moments in these processes, a definition of 
`normal' must be established. This paper proposes a new statistical 
model for such systems, modeled as dynamic networks, to address this challenge. It assumes that 
vertices fall into one of $k$ types and that the probability of edge formation at a particular
time depends on the types of the incident nodes and the current time. The time
dependencies are driven by unique seasonal processes, which many systems
exhibit (e.g., predictable spikes in geospatial or web traffic each day). The paper 
defines the model as a generative process and an inference procedure to recover 
the `normal' seasonal processes from data when they are unknown.
 An outline of anomaly detection experiments to be completed over
Enron emails and New York City taxi trips is presented.
\end{abstract}

%
%



\maketitle

\section{Introduction \& Motivation}

\begin{figure}
\centering
\includegraphics[width=.5\textwidth]{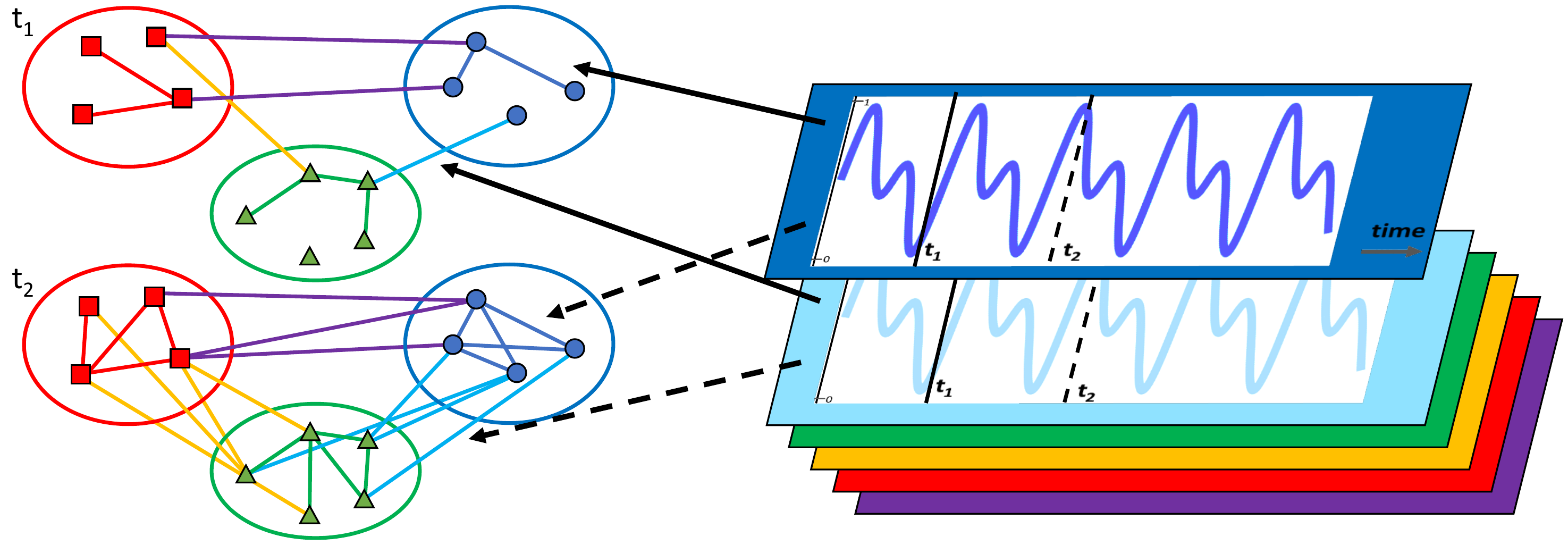}
\caption{Example of seasonality in dynamic networks. In this three class network, 
a latent seasonal process determines a probability of edge formation at times $t_1$ and $t_2$, subject
to both process and measurement noise. A different process (colored plates on the right) 
affect probabilities for edges connecting unique pairs of node types.
The paper presents a statistical model codifying these ideas, which may be useful in 
comparison, prediction, and anomaly detection tasks on dynamic complex systems.}
\label{fig:SDSBM-Overview}
\vspace{-15px}
\end{figure}

Many complex systems exhibit regular, time dependent, {\em seasonal} patterns. For example, human movement 
patterns are driven by the time of day~\cite{song2010limits}, and vehicle traffic densities exhibit 
predictable increases at certain hours causing rush hours and decreases at night~\cite{li2009temporal}. This same `seasonal', time dependent effect occurs when monitoring network bandwidth usage \cite{yoo2016time}, when counting the number of clicks per day on a web page\cite{gonccalves2008human}, or when tracking mobile traffic levels \cite{xu2016big}.

We look to bring the notion of seasonality to statistical network modeling with a new kind of dynamic stochastic block model (DSBM). A DSBM asserts
that system components (nodes) are grouped into several types, and the probability of 
observing a component relation or interaction (edges) are determined by 
the types of the incident nodes and time. Different kinds of DSBM 
consider different assumptions about the network formation 
process, 
but none consider seasonality. The model also explicitly introduces the notation of a \textit{measurement noise}, in addition to the existing randomness of the stochastic block model. The importance of this additional parameter is explored through experiments, and the significance on real datasets is hypothesized.

A conceptual overview of the model we propose is given in Figure~\ref{fig:SDSBM-Overview}.
It fuses structural time series (plates on the right of Figure~\ref{fig:SDSBM-Overview}) with a generative network model.
We call this a seasonal DSBM (SDSBM). This model was first introduced in \cite{robinson2017seasonality}. In this paper we expand upon that work, providing discussion of an additional variance parameter introduced into the state space model, details of the scalable fitting to data by Kalman Filters, parameter learning by expectation-maximization, and how the SDSBM can be used to define an anomaly detector.

The rest of the paper is organized as follows: section 2 discusses the related statistical dynamic network formulations and their limitations to seasonal data, section 3 specifies the generative details of the SDSBM, section 4 an inference and learning procedure using a Kalman Filter and expectation-maximization (EM) is created, section 5 defines the anomaly detector and section 6 discuss in progress current and future experiments. This manuscript will be updated as experiments are completed.


\section{Related Work}
The original stochastic block model (SBM) \cite{holland1983stochastic} has been a successful statistical random network. It has been theoretically explored \cite{karrer2011stochastic} and applied \cite{fortunato2010community} extensively on static networks, mainly for community detection. A survey of SBMs is available \cite{matias2014modeling}. 

With more access to large, time dependent datasets, the study of dynamic networks (i.e. a time ordered sequence of static networks) has become more popular. It is only natural with the success of the original SBM that dynamic variants would be created. In \cite{xu2013dynamic}, the authors assume a DSBM, where the probability of edge formation follows a random walk in time. Using an extended Kalman Filter augmented with a local search, a model fitting procedure is defined. In \cite{xu2015stochastic}, the same author extends the work to removes the hidden Markov assumptions on edge-level dynamics, allowing for dependency across time. In \cite{Yang2011}, instead of modeling changing probability of edge formation, the authors allow vertices to randomly change types through time. Using a Bayesian framework, posterior distributions of the model parameters are derived. Many other variants exist \cite{xing2010state, ho2011evolving}, each encoding slightly different assumptions on how the system evolves with time. None of these approaches explicitly model the dynamic behavior following time series techniques, and would be unable to effectively model patterns in the seasonal datasets described earlier.

Anomaly detection over dynamic networks is a popular problem. An excellent survey paper on the topic is available \cite{ranshous2015anomaly}. The survey authors create a taxonomy of different anomaly detection approaches, for anomalous vertices, anomalous edges, anomalous subgraphs, and change point detection. To quantify what is `anomalous', researchers has explored community detection, compression, decomposition, distance and probabilistic metrics. In this paper, the SDSBM can be used to defined a probabilistic anomaly detector for anomalous subgraphs and change point detection. We assume graphs are being generated from a seasonal time series, and significant deviation from the seasonal pattern is considered anomalous.

\section{Model Specification} \label{sec:modelSpec}
We first specify the model of the seasonal processes controlling edge dynamics. 
We assume that time is discrete, with the current time $t$ representing a 
time period of some resolution. For fixed datasets, $t \in [1,2, ...,T]$, where $T$ is the largest time step. We also assume the node types are provided.
For each pair of node types $a$ and $b$, we consider a structural
time series with a {\em bias} $m_t^{(a,b)}$ establishing an
anchor for values of the time series and a {\em seasonal offset} $s_t^{(a,b)}$ that 
shifts the bias by the current seasonality position. The process at time $t$, denoted $c^{(a,b)}_t$, is  
\begin{equation} \label{eq:genLevel}
c^{(a,b)}_t = m^{(a,b)}_t + s^{(a,b)}_t \end{equation}
with bias $m^{(a,b)}_t$ described by
$m^{(a,b)}_t = m^{(a,b)}_{t-1} + \delta_{m^{(a,b)}_t}$ 
where $\delta_{m^{(a,b)}_t} \sim \mathcal{N}(0,q^{(a,b)}_m)$ models possible \textit{process noise}. 
The set of seasonal offsets are stored in a vector
 $\s^{(a,b)} =(s^{(a,b)}_1, s^{(a,b)}_2, ..., s^{(a,b)}_d)$ having $d$ components. 
 $d$ reflects either the length or the resolution
of a seasonal process (e.g., $d=60$ to model per minute changes over a process that
cycles per hour) and is assumed to be provided by the user of the model. The components are: 
\begin{equation}\label{eq:genSeasonality}
s^{(a,b)}_t = -\sum_{i=1}^{d-1}s^{(a,b)}_{t-i} + \delta_{s^{(a,b)}_t} 
\end{equation}
where $\delta_{s^{(a,b)}_t} \sim \mathcal{N}(0,q^{(a,b)}_s)$. 
This form 
enforces a  zero-sum constraint to increase identifiability~\cite{murphy2012machine}. 
It should be emphasized that
$q^{(a,b)}_m$ and $q^{(a,b)}_s$ control the noise of the underlying seasonal process. This noise level affects the \textit{evolution} of the process across time. For low noise levels, the time series will closely follow a rigid seasonal pattern that repeats season and after season (i.e. closer to a fixed sine wave). For high noise levels, the times series will deviate from the seasonal pattern, becoming largely unpredictable (i.e. closer to a random walk). When learning from data, then using the inferred seasonal pattern for forecasting, a lower process noise is preferred for effective predictions.

To model how the seasonal processes govern the shape of a dynamic network, 
we define a random variable $e^{(a,b)}_t \in [0,1]$ as the {\em expected density} of edges spanning node types $(a,b)$ at time $t$:
\begin{equation}\label{eq:genDensity}
e^{(a,b)}_t = c_t^{(a,b)} +  \epsilon_{e^{(a,b)}_t}
\end{equation}
where $\epsilon_{e^{(a,b)}_t} \sim \mathcal{N}(0,r^{(a,b)})$ models possible
\textit{measurement noise} at time $t$. It is importance to understand how the process noise and measurement noise influence the model in different ways. The process noise will control the evolution \textit{through time}, while the measurement noise controls the variation at a single time step. Additional exploration and discussion of these noise parameters on synthetic data are completed in results section \ref{sec:experiments}.

Next we define an adjacency matrix $A_t$, where $[A_t]_{ij} = 1$ if there exists an edge between nodes $i$ and $j$ at time $t$ and $[A_t]_{ij}=0$ otherwise. 
Denote $A_t^{(a,b)}$ as the submatrix of $A_t$ only containing the rows and columns representing type $a$ and type $b$ nodes. Then $A_t^{(a,b)}$ is defined by the random variable:
\begin{equation} \label{eq:genAdj}
[A_t^{(a,b)}]_{ij} \sim Bernoulli(e_t^{(a,b)})
\end{equation}
Repeating this process for all blocks $(a,b)$ and time steps $t$ will generate a desired dynamic network $\mathcal{A} = \{A_1, A_2, ..., A_t\}$.

\section{Model Fitting}
We now describe an inference procedure to fit the model to an observed $\mathcal{A}$. 
Since seasonal processes are latent in data, the task is to estimate a posterior distribution 
on each $m^{(a,b)}_t$ and $s^{(a,b)}_t$ for each pair of node types $(a,b)$. 
Kalman filters~\cite{kalman1960new} are an appropriate tool for this task, but requires 
transforming the generative model into a state-space model (SSM). A SSM 
is a time series model with \textit{hidden state} and \textit{observed} variables~\cite{murphy2012machine}; 
here we define $\x_t$ as hidden state and $w_t$ as observed variables, respectively. We can notice the bias and seasonal offsets from before were hidden, while the adjacency matrix is observed, foreshadowing the structure to be defined. 
A SSM creates observations at time $t$ by two linear models: 
An {\bf observation model}  

\begin{equation}\label{eq:obsModel}
w_t = \h \x_{t} + \epsilon_t
\end{equation}
and a {\bf transition model} 

\begin{equation}\label{eq:transModel}
\x_t = \Gb \x_{t-1} + \Delta_t
\end{equation}

Here, observations $w_t$ are generated by a transformation of the output 
(defined by $\h$) of the underlying transition model. The transition model describes transformations within a hidden state space where transitions 
from time $t-1$ to time $t$ are defined by the matrix $\Gb$. Observations and transitions are
to be subject to time dependent random noise, which are modeled by gaussian distributions
$\epsilon_t \sim \mathcal{N}(0, b_t)$ and $\Delta_t \sim \mathcal{N}(0, \Q_t)$
With $b_t$ and $\Q_t$ controlling the amount of observation and transition noise, respectively.
Assuming parameters $\theta_t = \{\h, \Gb, b_t, \Q_t\}$ are known, a Kalman Filter can be used to derive the exact posterior $Pr(\x_t | w_1, w_2, ...,w_t ; \theta_t)$, i.e., the probability of the hidden state value at time $t$ given the observations up to and including time $t$~\cite{murphy2012machine}. 
 
Now we transform the model specification into a state space to define the transition 
model $\x_t = \Gb \x_{t-1} + \Delta_t$. As we are assuming edges of different vertex types $(a,b)$ are independent of each other, we will formulate the inference in terms of a pairing $(a,b)$. The full inference is completed by repeating the process for all pairs $(a,b)$.  
First we transform Equations~\ref{eq:genLevel} and~\ref{eq:genSeasonality} to define the 
hidden state variable $\x_t$ and state transition $\Gb$. The hidden state will be composed of the bias and vector of seasonal offsets as a $d \times 1$ seasonal state vector for a 
period of length $d$: 
\begin{equation}
\x^{(a,b)}_t = \begin{bmatrix}
    m^{(a,b)}_t&
    s^{(a,b)}_t&
    s^{(a,b)}_{t-1}&
    \hdots&
    s^{(a,b)}_{t-d+2}
\end{bmatrix}^T
\end{equation}
where $^T$ is the transpose operator. Note that all the seasonal offsets from $\s^{(a,b)}$ are maintained in the state for a given $t$, with the $d$th seasonal offset implicitly defined based on the zero-sum constraint. Now to perform the state transition from time $t-1$ to time $t$ we define a $d\times d$ matrix $\Gb$:

\begin{equation} \label{eq:gb}
\Gb =  \begin{bmatrix}
    1 & 0 & 0 & \dots & 0 & 0 \\
    0 & -1 & -1 & \dots & -1 & -1 \\
    0 & 1 & 0 & \dots & 0 & 0 \\
    0 & 0 & 1 & \dots & 0 & 0 \\
    \vdots & \vdots & \vdots & \ddots & \vdots & \vdots \\
    0 & 0 & 0 & \dots & 1 & 0
\end{bmatrix}
\end{equation}
In $\Gb$, we see that multiplication of the first row of $\Gb$ by $\x^{(a,b)}_t$ yields Equation~\ref{eq:genLevel} without noise being added, as only the bias term is updated.
Multiplication of the second row of $\Gb$ by $\x^{(a,b)}_t$ will update a \textit{single} seasonal offset as shown in Equation~\ref{eq:genSeasonality}. The remaining rows of $\Gb$ serve to permute the remaining seasonal offsets, such that each offset $s^{(a,b)}_i$ is updated after a full period of $d$ time steps. Each time step will update the most current seasonal offset, and shift the remaining offsets right one index in the state vector. Next we define the $d \times 1$ noise vector $\Delta_t = [ \delta_{m^{(a,b)}_t}, \delta_{s^{(a,b)}_t}, 0, \dots, 0 ]^T$ where the first element is the bias noise in Equation~\ref{eq:genLevel}, the second element is seasonal noise in Equation~\ref{eq:genSeasonality}, 
and the remaining elements are all 0 as there is no additional noise for permuting the seasonal offsets.  These noise values are sampled from a zero mean gaussian with $d \times d$ covariance matrix $\Q = \texttt{diag}[q_m^{(a,b)}, q_s^{(a,b)},0,\dots,0]$. Assuming the bias noise $\delta_{m^{(a,b)}_t}$ and seasonal offset noise $\delta_{s^{(a,b)}_t}$ are independent, the off-diagonal elements of $\Q$ are zero. The remaining elements are all zero, reflecting the lack of noise for the permutation operations. We can assume $\Q$ is stationary, and drop the dependence on $t$.  It is important to notice this is a \textit{degenerate} variance matrix as it is singular. It is a combination of stochastic transformations for equations for variance of bias $q_m^{(a,b)}$ (\ref{eq:genLevel}) and variance of seasonality $q_s^{(a,b)}$ (\ref{eq:genSeasonality}) and deterministic permutation of past seasonal offsets. When inferring this matrix via expecation-maximization, modification will be made to enforce the proposed form. This complete formulation of the transition model is not new, and has been completed by other researchers such as in \cite{davis2005modeling}. 


Our next task is to transform Equations~\ref{eq:genDensity} and~\ref{eq:genAdj} into the
observation model $w_t = \h \x_{t} + \epsilon_t$. To do this, we will need to define some additional 
variables, and take advantage of a result of the central limit theorem for a large number of vertices with types $(a,b)$
to create an approximate gaussian transformation. First we need a count of the number of \textit{possible edges} in block $(a,b)$, so if there are $|a|$ nodes of type $a$ and $|b|$ nodes of type $b$ then define:
\begin{equation}\label{eq:n}
   n^{(a,b)}=\begin{cases}
   a=b & \frac{|a|(|a|-1)}{2} \\
   a\neq b &  |a||b|\\
   \end{cases}
\end{equation}
Also define the random variable $p_t^{(a,b)} \sim Binomial(e^{(a,b)}_t, n^{(a,b)})$ as the number 
of {\em formed edges} 
in block $(a,b)$ at time $t$, where $e^{(a,b)}_t$ is the expected edge density as 
determined by Equation~\ref{eq:genDensity}. $p_t^{(a,b)}$ is simply a more mathematically convenient way to define the edge generation process from Equation~\ref{eq:genAdj} and does not change the overall model. Having a binomial random variable, for large enough $n^{(a,b)}e_t^{(a,b)}$ we can apply 
the central limit theorem to approximate the distribution of $p_t^{(a,b)}$ as gaussian:
 \begin{equation} \label{eq:sumEdges}
 	p_t^{(a,b)} = n^{(a,b)}e_t^{(a,b)} +  \omega_{p^{(a,b)}_t}
 \end{equation}
where $\omega_{p^{(a,b)}_t} \sim \mathcal{N}(0, u_t\bk)$ with $u_t\bk := n^{(a,b)}e_t^{(a,b)}(1-e_t^{(a,b)})$ is {\em observation noise} that is time dependent on $e_t^{(a,b)}$. This represents inherent randomness associated with a repeated binary decision process. The amount of uncertainty in the process is directly coupled with the probability $e_t^{(a,b)}$ of each individual decision. To give a more concrete example, in a geospatial context, a seasonal process may dictate that many people travel from home to work at 8am. In this example, locations of the geospace are system components (vertices), and movement between these locations are interactions (edges). When looking at a individual, there is a binary decision to go to work (and create an edge in the network), or stay at home (no edge formed), where a collection of many individual will create the repeated binary decision processes.  In the SDSBM, each system component will have a type, such that each individual home is a vertex and all these homes can have the same type of `residence'.

Now that we have successfully transformed our edge sampling procedure from Equation~\ref{eq:genAdj} to an approximately gaussian formulation in Equation~\ref{eq:sumEdges}, 
we can return to defining observation model parameters $w_t$ and $\h$. We will define the observed variable $w^{(a,b)}_t$ as the number of formed edges  $p_t^{(a,b)}$. To create $\h$, 
define a transformation which takes as input a seasonal state vector $\x_t^{(a,b)}$ and produces as output the number of formed edges $w^{(a,b)}_t$. By combining the operations of Equations~\ref{eq:genDensity} and~\ref{eq:sumEdges} we define:
\begin{equation}
\h = \begin{bmatrix}
    n^{(a,b)} & n^{(a,b)} & 0 & \dots & 0
\end{bmatrix}
\end{equation}
Examining $w^{(a,b)}_t = \h\x^{(a,b)}_t$ closer, we can see this multiplication both sums the bias $m_t^{(a,b)}$ and first seasonal offset $s^{(a,b)}_t$ following Equation~\ref{eq:genDensity}, 
and multiplies by $n^{(a,b)}$ to match the repeated Bernoulli trials from 
Equation~\ref{eq:sumEdges}. Finally, we model the variance of the 
noise $\epsilon_t$ by summing measurement noise
$\epsilon_{e_t^{(a,b)}}$ and observation noise $\omega_{p^{(a,b)}_t}$. Typical applications of the SSM only have the observation noise such as in \cite{xu2013dynamic}, where the inclusion of a measurement noise is a novel contribution of this paper. This definition 
allowing for more flexible modeling of many complex systems, and is demonstrated through synthetic results in section \ref{sec:experiments}. $\epsilon_t$ is sampled from a zero mean gaussian distribution with time dependent variance $b^{(a,b)}_t = u_t\bk + (n^{(a,b)})^2r^{(a,b)}$.

\begin{figure}
\centering
\includegraphics[width=.15\textwidth]{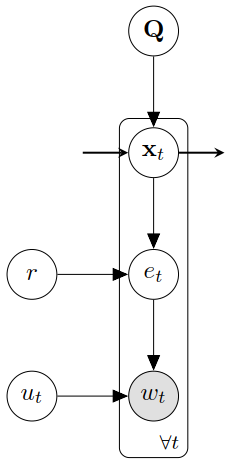}
\caption{Graphical model of the SDSBM in state space model form. The darker circles are assumed given, while the white circle are inferred.}
\label{fig:graphical-model}
\end{figure}

We have now transformed the generative procedure to the suitable SSM to allow easy inference via the Kalman Filter. Given an initial gaussian belief state $Pr(\x^{(a,b)}_0)$ with mean $\mub_0^{(a,b)}$ and variance $\Sigmab_0^{(a,b)}$, all subsequent belief states will be gaussian as well. A graphical model representation of the SDSBM is provided in figure \ref{fig:graphical-model}

\subsection{Kalman Filtering and Smoothing}

The recursive, closed-form updates for the posterior distribution $Pr(\x_t^{(a,b)} | w_1^{(a,b)},w_2^{(a,b)},...,w_t^{(a,b)} ; \theta_t^{(a,b)})$ follow from the Kalman Filter \cite{kalman1960new}. For the remainder of this section the explicit reference of blocks $(a,b)$ is dropped for notational simplicity. The same equations are repeated independently for all $(a,b)$. The formulation repeatedly applies two steps of \textit{prediction step} and \textit{update step}. Given the observations up to time $t-1$, the prediction step uses the assumed transition model in equation \ref{eq:transModel} to forecast the distribution at time $t$. That is:

\begin{equation}
Pr(\x_t | w_{1:t-1}; \theta_t) = \mathcal{N}(\mub_{t|t-1}, \Sigmab_{t|t-1})
\end{equation}
\begin{equation}\label{eq:muPred}
\mub_{t|t-1} = \Gb\mub_{t-1|t-1}
\end{equation}
\begin{equation}\label{eq:sigmaPred}
\Sigmab_{t|t-1} = \Gb\Sigmab_{t-1|t-1}\Gb^T + \Q
\end{equation}

where $\mub_{t|t-1} := \mathbb{E}[\x_t | w_{1:t-1}]$ and $\Sigmab_{t|t-1} := \mathbb{E}[(\x_t - \mub_{t|t-1})(\x_t - \mub_{t|t-1})^T| w_{1:t-1}]$. Now given the observation at time $t$, the update step combines the predicted state with the new observation to refine the estimate. That is:

\begin{equation}
Pr(\x_t | w_{1:t} ; \theta_t) = \mathcal{N}(\mub_{t|t},\Sigmab_{t|t})
\end{equation}
\begin{equation}\label{eq:muFilter}
\mub_{t|t} = \mub_{t|t-1} + \mathbf{k}_t(w_t - \h \mub_{t|t-1})
\end{equation}
\begin{equation}
\Sigmab_{t|t} = (\mathbf{I} - \mathbf{k}_t\h)\Sigmab_{t|t-1}
\end{equation}
\begin{equation}
\mathbf{k}_t =\Sigmab_{t|t-1}\h^T ( \h \Sigmab_{t|t-1} \h^T + \mathbf{b}_t)^{-1}
\end{equation}


where $\mub_{t} := \mathbb{E}[\x_t | w_{1:t}]$ and $\Sigmab_{t|t} := \mathbb{E}[(\x_t - \mub_{t|t})(\x_t - \mub_{t|t})^T| w_{1:t}]$, and $\mathbf{k}_t$ is known as the Kalman gain matrix. To build some intuition on these equations, we can look at equation \ref{eq:muFilter} closer. This equation calculates the new mean as a combination of the predicted mean plus a correction factor of the residual $w_t - \h \mub_{t|t-1}$, scaled by the Kalman gain $\mathbf{k}_t$. If $|\mathbf{k}_t|$ is large, which will occur if the ratio of variance parameters $\frac{\Q}{b_t}$ is large, then the new mean will be largely estimated based on the current observation $w_t$. If $|\mathbf{k}_t|$ is small, which will occur if the ratio of $\frac{\Q_t}{b_t}$ is small, the new mean will be mainly estimated based on the prediction $\mub_{t|t-1}$. This formulation can emphasize the impact $\Q$ and $b_t$ on the state estimates. Starting at $t=0$ with an initial guess of $\mu_0$ and $\Sigma_0$ for equations (\ref{eq:muPred}) and (\ref{eq:sigmaPred}), posteriors can be recursively calculated using equations the prediction step followed by the update step until $t=T$. 

For offline problems where all data $w_1,...,w_T$ have been collected before inference, we can use the \textit{Kalman Smoother} to improve the state estimates. The Kalman Filter only assumes knowledge of \textit{past} observations to estimate the distribution at time $t$, while the Kalman Smoother assumes knowledge of \textit{past and future} observations. The Kalman Smoother equations are given as:

\begin{equation}
Pr(\x_t | w_{1:T}; \theta_t) = \mathcal{N}(\mub_{t|T}, \Sigmab_{t|T})
\end{equation}
\begin{equation}
\mub_{t|T} =  \mub_{t|t} + \Sigmab_{t|t}\Gb^T\Sigmab_{t+1|t}^{-1}(\mub_{t+1|T} - \mub_{t+1|t})
\end{equation}
\begin{equation}
\Sigmab_{t|T} =  \Sigmab_{t|t} + \Sigmab_{t|t}\Gb^T\Sigmab_{t+1|t}^{-1}(\Sigmab_{t+1|T} - \Sigmab_{t+1|t})(\Sigmab_{t|t}\Gb^T\Sigmab_{t+1|t}^{-1})^T
\end{equation}

where $\mub_{t|T} := \mathbb{E}[\x_t | w_{1:T}]$ and \\ $\Sigmab_{t|T} := \mathbb{E}[(\x_t - \mub_{t|T}) (\x_t - \mub_{t|T})^T | w_{1:T}]$. By taking into account the additional information of future estimates, we will more accurately recover the true mean and variance at each time step. Starting with $t=T$, using the final Kalman Filter estimate of $\mub_{T|T}$ and $\Sigmab_{T|T}$ from before, recursively update the smoothed estimates until $t=0$. Now the full posterior distribution for each time step can be inferred. These inference algorithms assumed knowledge of $\Theta = \{\theta_1, \theta_2, ..., \theta_T\} = \\ \{u_1,u_2,...,u_T, r, \Q, \mub_0, \Sigmab_0 \}$ a priori. In the next section we discuss how to estimate $\Theta$.

\subsection{Expectation-Maximization}

In most applications, $\Theta$ must be inferred from the data. To estimate each $\{u_1, u_2, ..., u_T\}$ we follow a necessary condition from the gaussian approximation from Equation \ref{eq:sumEdges}. Using the prediction step of the Kalman Filter in Equation \ref{eq:muPred}, we estimate each $u_t$ as:

\begin{equation}
u_t = \h\mub_{t|t-1}(1 - \frac{\h\mub_{t|t-1}}{n})
\end{equation}

This is similarly done in existing DSBM from \cite{xu2013dynamic}. This leaves us to learn the time-invariant parameters of $\phi := \{r, \Q, \mub_0, \Sigmab_0\}$. The SSM offers a natural form for learning with the expectation-maximization (EM) algorithm to get locally optimal point estimates. EM is suitable for problems where the model depends on one the parameters of statistical models, and two on latent variables. In the SSM, the $\Theta$ for all $t$ are the statistical parameters, and the $\mub_{t|T}$ and $\Sigmab_{t|T}$ are the latent variables (or $\mub_{t|t}$ and $\Sigmab_{t|t}$ if using just Kalman Filter). The setup for EM on the standard SSM is discussed in detail in \cite{bishop2007pattern}. We will now define the steps of the EM algorithm, with the addition of the new variance term defined in equation \ref{eq:genDensity}. Let us denote the estimated parameters of the $ith$ iteration of EM as $\phi^{i}$. Initial guess from $\phi^0$ are provided by the user and can be default set to 1 if no domain context is available. EM works in two steps of \textit{expectation step} followed by \textit{maximization step}, then iterating repeated between the two until convergence.

First define the full log likelihood using the SSM transition and observation equations:
\begin{equation}
\begin{split}
\ln Pr(\x_{0:T}, w_{1:T} | \Theta) = \\
\ln \mathcal{N}(\mub_0, \Sigmab_0) + \\
\sum_{t=1}^T\ln \mathcal{N}(\Gb \x_{t-1},\Q) + \\
\sum_{t=1}^T \ln \mathcal{N}(\h \x_{t},u_t + n^2 r))
\end{split}
\end{equation}

The elegance of using EM with SSM becomes clear as the $\phi$ parameters are separately nicely between the three log likelihood terms. Now in the expectation step, calculate the expected value of the log likelihood function with respect to the conditional distribution:

\begin{equation}\label{expectation}
Q(\phi, \phi^i) = \mathbb{E}_{\x_{0:T}|\phi^i}[\ln{p(w_{1:T}, \x_{0:T}; \phi)}]
\end{equation}

Using the Kalman Filter and Kalman Smoother equations defined earlier, we recursively estimate the latent variables. In the end, collect results of:

\begin{equation}
\mathbb{E}[\x_t \x_{t-1}^T] = J_{t-1} \Sigma_{t|T} + \mu_{t|T}\mu_{t-1|T}^T
\end{equation}
\begin{equation}
\mathbb{E}[\x_t \x_t^T] = \Sigma_{t|T} + \mu_{t|T}\mu_{t|T}^T 
\end{equation}
\begin{equation}
J_t := \Sigmab_{t|t}\Gb^T\Sigmab_{t+1|t}^{-1}
\end{equation}

This completes the expectation step. Now in the maximization step, find the $\phi$ parameters which maximize this log likelihood. More formally:

\begin{equation}
\phi^{i+1} = \argmax_{\phi} Q(\phi , \phi^{i})
\end{equation}

As the $\phi$ parameters are well separated in the likelihood, the maximization derivations can be completed separately for each term. For the initial guesses $\mub_0$ and $\Sigmab_0$, the simple updates are:
\begin{equation}
\mub_0^{i} = \mub_{0|T}
\end{equation}
\begin{equation}
\Sigmab^{i}_0 = \Sigmab_{0|T}
\end{equation}

For the observation variance $r$, the parameter is set by finding r which maximizes function:

\begin{align*}
\sum_{t=1}^T\ln(\mathcal{N}(\h\x_{t},u_{t} + n^2r)) = \\
-\sum_{t=1}^T\frac{\ln|u_{t|t-1} + n^2r|}{2} \\
-\frac{1}{2}\sum_{t=1}^T\frac{(w_tw_t^T - w_t \mathbb{E}[\x_t]^T(\h)^T - \h \mathbb{E}[\x_t]w_t^T + \h \mathbb{E}[\x_t\x_{t}^T](\h)^T)}{(u_{t|t-1} + n^2r)}
\end{align*}

This task can be completed using standard optimization routines such as gradient descent.

For the final variance matrix $\Q$, some modifications need to be made due to not being full rank (only the first two diagonal elements are nonzero). This approach is originally described in \cite{harvey1990estimation}. Augment the state vector with two additional elements such that $\x^*_t := [\x_t^T, s_{t-d},m_{t-1}]^T$. To remain valid in the original SSM formulation, modifications also need to be made for $\Gb$, $\h$, and $\Q$.

\begin{equation}
\Gb^* = \left[
\begin{array}{cccccc|cc}
    1 & 0 & 0 & \dots & 0 & 0  & 0 & 0\\
    0 & -1 & -1 & \dots & -1 & -1 & 0 & 0\\
    0 & 1 & 0 & \dots & 0 & 0 & 0 & 0\\
    0 & 0 & 1 & \dots & 0 & 0 & 0 & 0\\
    \vdots & \vdots & \vdots & \ddots & \vdots & \vdots & \vdots & \vdots \\
    0 & 0 & 0 & \dots & 1 & 0 & 0 & 0 \\ \hline
    0 & -1 & -1 & \dots & -1 & -1 & 0 & 0 \\
    1 & 0 & 0 & \dots & 0 & 0  & 0 & 0
\end{array}
\right]
\end{equation}


For $\h^*$ append two additional zeros, and for $\Q^*$, also append additional zeros. This two modification have no effect on the computation, and are just necessary for correct dimensions. 


Finally with the the augmented state vector, we can analytically find the $\Q^*$ which maximizes $\sum_{t=1}^T\ln \mathcal{N}(\Gb \x_{t-1},\Q)$, with solutions:

\begin{equation}
q_m = \frac{(d_1\mu^*_{t|T})^2 + d_1\sum_{t=1}^T \Sigma^*_{t|T} d_1^T}{T}
\end{equation}
\begin{equation}
q_s = \frac{(d_2\mu^*_{t|T})^2 + d_2\sum_{t=1}^T \Sigma^*_{t|T} d_2^T}{T}
\end{equation}

where $d_1 = [1,0,\dots,0,-1]$ and $d_2 = [0,1, 0, \dots, 0, -1, 0]$.

Now if we iteratively perform the expectation and maximization steps, we will converge to a locally optimal solution for $\phi$. Given network data $\mathcal{A} = \{A_1, A_2, ..., A_t\}$, vertex types, length of seasonality $d$, and initial guesses for $\Q$, $r$, $\mub_0$, and $\Sigmab_0$, the seasonality of a dynamic network will be extracted.



\section{Anomaly Detection}


One of the applications of the created seasonal DSBM is for anomaly detection. An SDSBM fit to data learns the `normal' or expected behavior of a system, and can then be used to define what is abnormal. Given a dynamic network $\mathbb{G}$ as a sequence of static networks $G_t$, $\mathbb{G} = \{G_1, G_2, ..., G_T\}$, anomalous graphs $G_t$ can be defined as the $\{G_t \in \mathbb{G} | \ln{L(G_t)} \leq c_0)\}$, for some threshold $c_0$, where $L$ is the log likelihood function for the dynamic network.

For the seasonal dynamic stochastic block model, the log likelihood of a graph at a time step is defined as the sum of the log likelihoods of each of the blocks. That is: 

\begin{equation}
\begin{split}
L(G_t) = \sum_{\forall{ab}}\ln Pr(w_t\bk | w_{1:T}\bk, \theta_t) =  \\
\sum_{\forall{ab}}\ln \mathcal{N}(\h\mub_{t|T}\bk, \h\Sigmab_{t|T}\bk\h^T + (u_t\bk + (n\bk)^2 r\bk))
\end{split}
\end{equation}

where $\mathcal{N}()$ is the continuous density function for the gaussian distribution.

This definition of an anomaly provides some interesting features. First as the definition of a graph likelihood is proportional to subgraph likelihoods based on node types $(a,b)$, this allows us to view anomalies at different levels. At the highest level, there are graph level anomalies. When looking at a single subgraph $(a,b)$, there are subgraph level anomalies. Whenever there is a graph level anomaly, it can be further investigated to see which subgraphs were most anomalous.

There are several methods to determine a reasonable threshold $c_0$. Ideally in cases where a dataset has labeled anomalies, experiments can be completed in the domain, establishing true positive and false positive rates $c_0$ is varied. Unfortunately in most cases, the existence of `true' anomalies is unknown and the former method will not be feasible. 
In those cases, we can take advantage of the probabilistic definition of the model. Depending on the users desired sensitivity of the detector, a threshold can be set using standard deviations of the gaussian distribution for each block. So using 3 standard deviations ($99.7\%$ of the distribution), the chance of an event occurring naturally outside of these bounds are approximately 1 in 370. This can give the user an intuitive notion of how unlikely a graph needs to be in order to be anomalous.


\section{EXPERIMENTS} \label{sec:experiments}

Synthetic experiments are completed to explore the impact of adding the additional \textit{measurement noise} parameter from Equation \ref{eq:genDensity}. The state space model as defined in Equations \ref{eq:obsModel} and \ref{eq:transModel} has two variance parameters, $\Q$ and $b_t$. $\Q$ contains $\{q_m, q_s\}$ which together define the level of process noise which controls the noise in the latent seasonality. $b_t$ in our definition is defined as the sum of $u_t$ the observation noise inherently created and controlled by the probability of the binary random decision process, and $r$ the level of measurement noise in the probability of the decision process. Existing applications of the state space model such as in \cite{xu2013dynamic} do not have the measurement noise (so $r = 0$). 

\begin{figure}
\centering
\includegraphics[width=.5\textwidth]{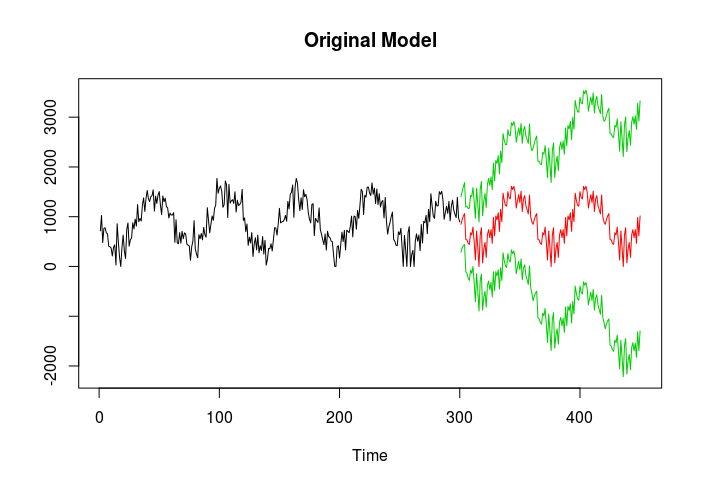}
\caption{This is a plot of data generated and fit from the typical state space model with only the process noise and observation noise, without the additional measurement noise parameter. The time series of edge counts data is in black, forecasted mean fit in red, and 95\% confidence bounds in green.}
\label{fig:original-model-var}
\end{figure}

To understand how this parameter will affect forecasts in the SSM, we setup a simple synthetic experiment. Data is generated using the state space model formulation with a small $\Q$, and the level of noise to be expected in real data (a `medium' amount) for $r$. First in figure \ref{fig:original-model-var}, the SDSBM is fit using the Kalman Filter and EM with $r$ being fixed to zero. This original form of the model will have to place the synthetic measurement noise into the optimization of $\Q$. This will results in a larger $\Q$ than is reasonable. When forecasting into the future, the large $\Q$ results is rapidly large confidence bounds. Given that the original data never became close to value 3000, it is not useful to have confidence bounds that wide. The forecasting is essentially useless. 

\begin{figure}
\centering
\includegraphics[width=.5\textwidth]{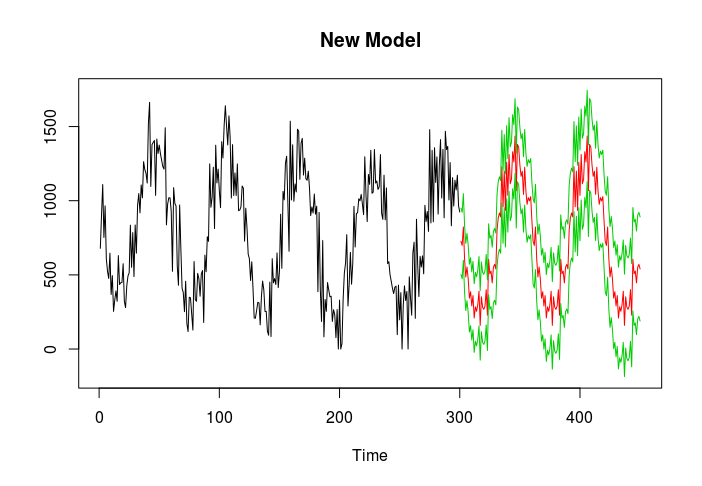}
\caption{This is a plot of data generated and fit from the new state space model with all three of the process noise, observation noise, and measurement noise. The time series of edge counts data is in black, forecasted mean fit in red, and 95\% confidence bounds in green.}
\label{fig:new-model-var}
\end{figure}

Now if we complete the same fitting, this time with a nonzero $r$, as shown in figure \ref{fig:new-model-var}. The model is able to keep the $\Q$ variance low, while capturing the additional variation in $r$. This effect is clearly seen in the confidence bounds of the forecast. The smaller bounds result in much more useful forecasting.

The main difference in $\Q$ and $r$ in forecasting, is the uncertainty due to $\Q$ cumulates with time, while uncertainty due to $r$ only occurs at the specific time step $t$. If a forecasting is predicting $y$ time steps into the future, the confidence bounds will be proportional to $y*\Q + r$.

In order for these result to be useful, we need to argue the noise from $r$ will exist and be significant in real dataset. The original problem this model has been developed for is in a surveillance application. Data of movement in a city is collected from a single wide area sensor. In this complex system, there is process noise of the seasonal process as people follow the daily patterns, there is measurement noise from the sensor itself, and there is the observation noise from the uncertainty in individuals' movements. Without the additional parameter, the uncertainty due to the sensor would results poor forecasts.

\subsection{Work in Progress}

Experiments on real datasets of Enron email network \footnote{Enron dataset available: https://www.cs.cmu.edu/~enron/} and NYC Taxi trips  \footnote{NYC Taxi dataset available: http://www.nyc.gov/html/tlc/html/about/trip\_record\_data.shtml} are ongoing, and will be added in a future version of this manuscript. For Enron, dynamic network can be created from daily email communication, where vertices are email users, and vertex types are based on the available job titles. This will result in a seasonal pattern over the week, as people are much more likely to send emails during the week, than on the weekends. By fitting the SDSBM and using our anomaly detector, it will be possible to detect graph level anomalies, which can be further explained by looking at the subgraphs of vertices of types $(a,b)$. It will be possible to search the email contents for an explanation behind the anomaly.

In the NYC taxi dataset, dynamic networks can by having pick up and drop off locations be vertices, and edges if a taxi trip is made in the time window. The vertex types can be selected based on the six boroughs of New York City. There is a clear seasonal pattern over a weekly period. Monday through Friday have a strong pattern based around the typical work and social schedules (spikes in the morning and evening), and Saturday and Sunday have their own seasonal pattern. Again, by fitting the data to the SDSBM and using the anomaly detector, interesting insights can be found. It is hypothesized significant events such as holidays, special events, and the weather will have a large impact on the network. A significant deviation from the seasonal taxi pattern will generally only occur due to these anomalous events.





\begin{acks}
This work is supported by industry and government partners at the National Science Foundation's I/UCRC Center for Surveillance Research and the Air Force Research Laboratory.
\end{acks}

\bibliographystyle{ACM-Reference-Format}
\bibliography{SDSBM-ref} 

\end{document}